\documentclass[12pt]{article}
\usepackage{graphicx}
\usepackage{subfig}
\usepackage{lineno}
\usepackage{hyperref}

\newcommand\pubnumber{ATL-PHYS-PROC-2014-247}
\newcommand\pubdate{\today}

\def\address{
On behalf of the ATLAS and CMS Collaborations\\
\vspace{0.5cm}
Sub-department of Particle Physics, University of Oxford,\\
 Keble Road, Oxford, OX1 3RH, United Kingdom\\
}

\def\Title#1{\begin{center} {\Large #1 } \end{center}}
\def\Author#1{\begin{center}{ \sc #1} \end{center}}
\def\Address#1{\begin{center}{ \it #1} \end{center}}

\newcommand\pubblock{\rightline{\begin{tabular}{l} \pubnumber\\
         \pubdate  \end{tabular}}}
\newenvironment{Abstract}{\begin{quotation}  }{\end{quotation}}
\newenvironment{Presented}{\begin{quotation} \begin{center} 
             PRESENTED AT\end{center}\bigskip 
      \begin{center}\begin{large}}{\end{large}\end{center} \end{quotation}}
\def\Acknowledgements{\bigskip  \bigskip \begin{center} \begin{large}
             \bf ACKNOWLEDGEMENTS \end{large}\end{center}
For fruitful discussions and comments on both the presentation and these proceedings, I would like to thank
N. Gutierrez, C. Heidemann, M. Martinez, A. Schmidt, F. Spano, J. Tseng, S. Willoq and J. Zhong. 
I gratefully acknowledge support from the Rhodes Trust. 
I would further like to thank the C R Barber Trust Fund as well as the conference organisers for funding this travel.
}
\begin{document}
% \linenumbers
\begin{titlepage}
\hspace{-1.4cm}\pubblock

\vfill
\Title{Searches with Boosted Objects}
\vfill
\Author{ Katharina Behr}%\support}
\Address{\address}
\vfill
\begin{Abstract}
\textit{Boosted objects} - particles whose transverse momentum is greater than twice their mass - are becoming increasingly important as the LHC
continues to explore energies in the TeV range. The sensitivity of searches for new phenomena beyond the Standard Model depends critically on the efficient reconstruction and
identification (``tagging'') of their unique detector signatures. This contribution provides a review of searches for new physics 
carried out by the ATLAS and CMS experiments that rely on the reconstruction and identification of boosted top quarks as well as boosted $W$, $Z$ 
and Higgs bosons. 
A particular emphasis is placed on the different substructure techniques and tagging algorithms for top quarks and bosons employed by the 
two experiments.
\end{Abstract}
\vfill
\begin{Presented}
XXXIV Physics in Collision Symposium \\
Bloomington, Indiana,  September 16--20, 2014
\end{Presented}
\vfill
\end{titlepage}
\def\thefootnote{\fnsymbol{footnote}}
\setcounter{footnote}{0}

\section{Introduction}
Despite its tremendous success - once again impressively demonstrated by the discovery of the long-predicted Higgs Boson in 2012 - 
the Standard Model (SM) of particle physics
is widely considered an incomplete theory. For one, it cannot explain the fact that the mass of the Higgs boson is light (hierarchy problem) 
nor does it offer a candidate for dark matter or satisfactorily explain the matter-antimatter asymmetry in the observed universe. 
Hence a number of extensions to the SM have been proposed whose predictions are currently under scrutiny by the LHC experiments.
These extensions include theories with warped extra-dimensions (Randall-Sundrum models), new strong interactions 
(Technicolour and others), an additional quark generation or vector-like quarks as well as supersymmetry. 
(For details of the specific models see the references in Sections~\ref{sec:BoostedTops}, ~\ref{sec:BoostedBosons},
~\ref{sec:TopPartners} and references therein.) 
Many of these models predict the existence of new heavy particles with large branching fractions into top quarks, 
heavy gauge bosons or the Higgs boson.
If these new states are sufficiently heavy their decay products are likely to have transverse momenta exceeding twice their rest masses.
These decay products are called \textit{boosted objects}.

The sensitivity of searches for new phenomena at high energies depends critically on the efficient reconstruction and identification
of boosted object decays. Boosted techniques first were applied in searches at the Tevatron (see~\cite{Altheimer:2013yza} for a recent review)
and developed into a fast %~\cite{Aaltonen:2014yja}
growing field of research during Run~I (2010-2012) of the LHC, as its higher center-of-mass energies of 7~TeV (2011) and 8~TeV (2012)
allow for abundant production of boosted objects across many final states.
This enabled the experiments to push the exclusion limits for many new particles into the TeV regime.\\
This document provides a review of the most recent searches with boosted objects carried out by the ATLAS~\cite{Aad:2008zzm} and 
CMS~\cite{Chatrchyan:2008aa} experiments
and presents some of the most commonly used reconstruction and identification techniques.
The rapid growth of the field makes it impossible to cover every single technique within the scope of this document. 
More details can be found in the proceedings of the BOOST workshop~\cite{Altheimer:2013yza}.

\section{Large-R Jets and Substructure}

\begin{figure}
\centering
\subfloat[]{
  \includegraphics[width = 0.50\textwidth]{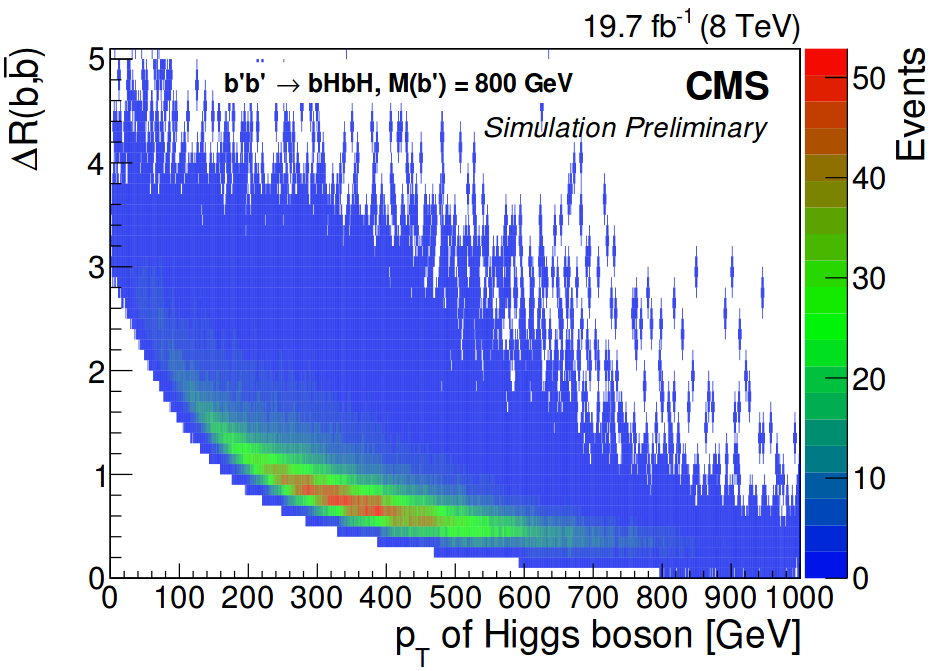}
}
\subfloat[]{
  \includegraphics[width = 0.39\textwidth]{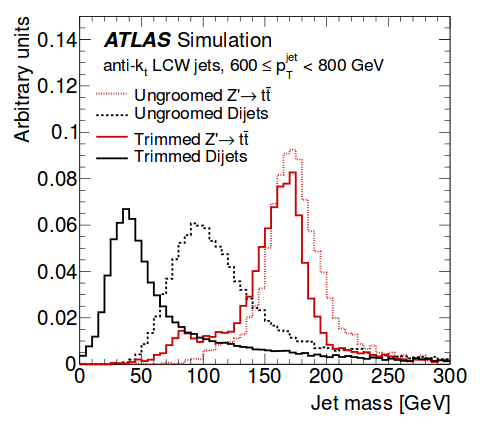}
}
\caption{(a) Angular separation between the two $b$ quarks from the decay of a Higgs boson as a function of $p_T^{H}$~\cite{CMS:2014afa}. 
(b) Mass distribution of the leading-$p_T^{jet}$ jet for ungroomed and trimmed jets on a $Z'\rightarrow t\bar{t}$ signal 
and a dijet background sample~\cite{Aad:2013gja}.}
\label{fig:dRSeparation}
\end{figure}

The defining property of boosted object decays is the fact that their decay products
appear collimated in the momentum direction of the boosted mother particle in the rest frame of the detector. Their angular separation
$\Delta R$ is inversely proportional to the transverse momentum $p_T$ of the mother particle with mass $m$ according to a simple rule of 
thumb:\footnote{
$\Delta R = \sqrt{\Delta \eta^2 + \Delta \phi^2}$ with pseudorapidity $\eta$ = -ln tan($\theta$/2). $\theta$ ($\phi$)
is the polar (azimuthal) angle of the ATLAS/CMS standard coordinate system, a 
right-handed orthogonal system with the z-axis tangential to the
beam pipe and the nominal interaction point in the detector centre as its origin.
%Given the cylindrical detector geometry, it is convenient to use of cylindrical coordinates (r, $\phi$) in the transverse plane.
} 
$\Delta R \approx 2 m/ p_T$. Figure~\ref{fig:dRSeparation}(a) illustrates this for the boosted decay $H \rightarrow b \bar{b}$.
Consequently, at high $p_T$, the decay products of a hadronically decaying object merge into a single large-$R$ jet 
with a characteristic substructure that allows one to distinguish these jets from those initiated by a single parton.\\
While large-$R$ jets can be reconstructed using any of the three common sequential recombination algorithms --- anti-$k_T$~\cite{Cacciari:2008gp}, 
Cambridge-Aachen (C-A)~\cite{Dokshitzer:1997in,Wobisch:1998wt} or $k_T$~\cite{Ellis:1993tq,Catani:1993hr} --- only the last two are suited for a
substructure analysis: they start by clustering close-by (C-A) or close-by and soft particles ($k_T$), effectively reversing the ordering of the 
parton shower. By undoing the last clustering step(s) and analysing properties of the subjets such as their relative $p_T$ fraction or 
angular separation, the presence of hard splittings in the jet can be probed. 
ATLAS and CMS use a number of different substructure variables to tag boosted object jets, the most prominent among which are the $k_T$ 
\textit{splitting scales} (see for example~\cite{TheATLAScollaboration:2013qia} and~\cite{CMS:2014fya})
and \textit{n-subjettiness}~\cite{Thaler:2010tr}
which measures how compatible the jet structure is with the ``n subjets'' hypothesis.

At high-luminosity hadron colliders, a major obstacle for analyses relying on large-$R$ jets is the presence of pile-up and the Underlying Event, 
both of which lead to soft, wide-angle contaminations that dilute the jet substructure. 
Various \textit{grooming} techniques that remove these contaminations have been developed, the most widely used of which are 
\textit{trimming}~\cite{Krohn:2009th}, \textit{pruning}~\cite{Ellis:2009su} and \textit{filtering}~\cite{Butterworth:2008iy}.
Figure~\ref{fig:dRSeparation}(b) illustrates the effect of trimming on the jet mass, a variable
widely used for boosted top identification. The trimmed distributions exhibit a significantly improved separation
between signal and background compared to the ungroomed case.

The efficient identification of jets originating from $b$-quarks is another crucial aspect for new physics searches. Dedicated performance 
studies targeted particularly at boosted topologies have been published~\cite{CMS:2013vea}.

\section{Searches with Boosted Top Quarks}\label{sec:BoostedTops}

Both ATLAS and CMS have conducted a wide range of searches in final states with boosted top quarks which rely on different combinations
of substructure variables and grooming techniques to efficiently identify hadronic decays of boosted top quarks. Many such dedicated
``top taggers'' have been developed and optimised for different final states and kinematic regimes. A review of these techniques can be found
in the performance notes by ATLAS~\cite{TheATLAScollaboration:2013qia} and CMS~\cite{CMS:2014fya}.

% Final states with boosted top quarks play an important role in searches for heavy particles decaying to $t\bar{t}$ pairs (\textit{resonances}), 
% heavy partners of the $W$ boson ($W'$) as well as supersymmetric top partners (\textit{stop quarks}).
% Searches for top-partners such as vector-like quarks which rely not only on boosted top quarks but also boosted gauge and Higgs bosons will be
% discussed separately in Section~\ref{sec:TopPartners}.

\subsection{$t\bar{t}$ Resonance Searches}
Searches for heavy resonances decaying into $t\bar{t}$ pairs have traditionally been the flagship applications for boosted techniques, 
both at the Tevatron and the LHC. ATLAS and CMS have published results in the single-lepton+jets (1$\ell+$jets) and 
the all-hadronic decay channels which rely on the reconstruction of hadronic decays of boosted top quarks. 
Two benchmark models, a leptophobic top-colour $Z'$ boson (narrow resonance) and a Kaluza-Klein gluon $g_{KK}$ arising in Randall-Sundrum (RS) models
(wide resonance), have been considered. 
For the sake of brevity, only results for the $Z'$ model will be discussed here.

\begin{figure}
\centering
\subfloat[]{
  \includegraphics[width = 0.49\textwidth]{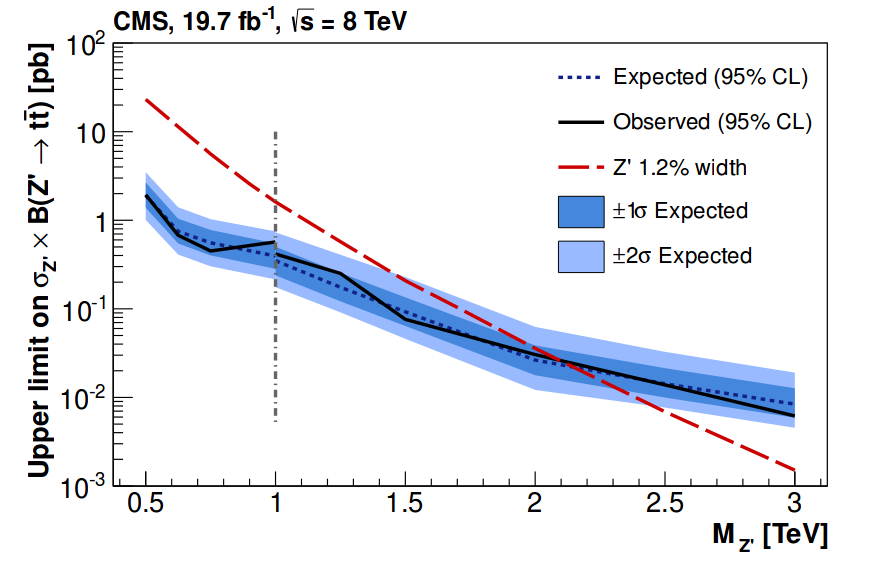}
}
\subfloat[]{
  \includegraphics[width = 0.45\textwidth]{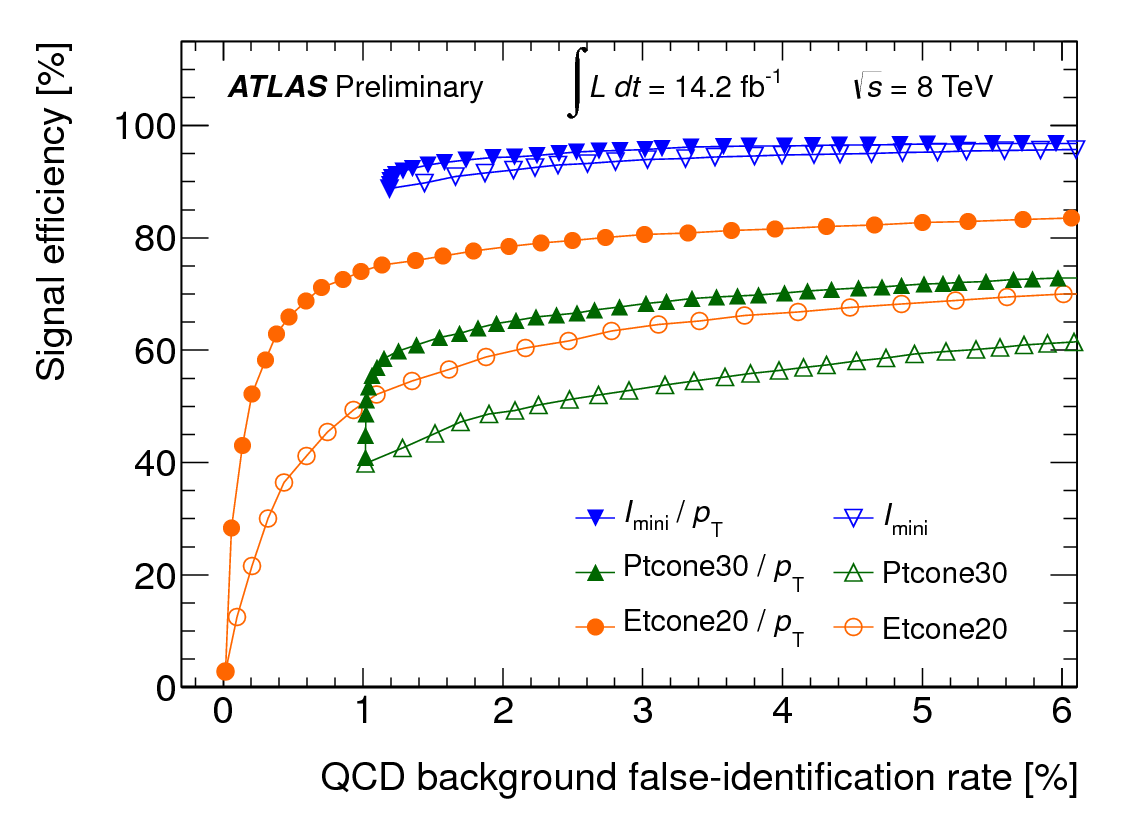}
}
\caption{(a) Upper cross-section limits on the production of $Z'$ using the combination of the semileptonic and all-hadronic channels. Above 1~TeV
the sensitivity is driven by the boosted channel~\cite{Chatrchyan:2013lca}.
(b) Performance of mini-isolation (blue curves) for highly energetic leptons compared to traditional fixed-cone isolation concepts~\cite{ATLAS:2012bpa}.}
\label{fig:ttbarRes}
\end{figure}

CMS has published results from the combination of the semileptonic and the all-hadronic decay channels 
using 19.7~$\textrm{fb}^{-1}$ of $\sqrt{s}$=8~TeV $pp$ collision data~\cite{Chatrchyan:2013lca}. 
In the semileptonic channel both a traditional resolved selection 
which relies on the reconstruction of the individual decay products of the top quarks
and a selection relying on boosted techniques are used: while the resolved selection requires exactly one isolated electron or muon as well as
$\geq4$ anti-$k_T$ jets with $R=0.5$, the boosted selection requires $\geq2$ anti-$k_T$ jets with $R=0.5$ and high transverse momentum.
The lepton isolation requirement is dropped completely to account for the fact that in boosted leptonic top quark decays the distance between the
lepton and the $b$-jet is small on average. The boosted channel dominates the sensitivity of the expected upper limit for $m_{t\bar{t}}>1$~TeV
as indicated in Figure~\ref{fig:ttbarRes}(a).

In the all-hadronic channel more sophisticated substructure techniques are needed to reduce the large multi-jet background. At least two C-A $R=0.8$ 
jets tagged by the \textit{CMSTopTagger}~\cite{CMS:2014fya} are required. This top tagger analyses the subjets and uses various kinematic criteria
such as a $W$ and top mass window requirement to identify the three-pronged substructure compatible with the decay 
$t \rightarrow Wb \rightarrow q\bar{q'}b$. The combination of all channels yields a 95\% CL lower limit on $m_{Z'}$ of 2.1 TeV.

ATLAS has published two separate searches in the semileptonic (resolved and boosted) and all-hadronic (boosted only) decay channels.
The semileptonic search~\cite{ATLAS:2012bpa} uses 14~$\textrm{fb}^{-1}$ of $\sqrt{s}$=8~TeV data. In the boosted channel the hadronically decaying top quark 
is reconstructed as a trimmed anti-$k_T$ 1.0 jet passing identification cuts on the jet mass, see Figure~\ref{fig:dRSeparation}(b),
and the $k_T$ splitting scale $\sqrt{d_{12}}$. 
In addition, this analysis uses a novel lepton isolation concept, \textit{mini-isolation}, which replaces the traditional
fixed-R isolation cone by one that shrinks inversely with increasing lepton transverse momentum $p_T^{lep}$, 
thus taking into account the collimation of the decay products of
boosted particles. Mini-isolation provides consistently high signal efficiency over the whole $p_T^{lep}$ range
and outperforms traditional isolation concepts for higher $p_T^{lep}$ as shown in Figure~\ref{fig:ttbarRes}(b).
The 95\% CL lower limit on $m_{Z'}$  resulting from the combination of the boosted and resolved channels is 1.7~TeV.

The ATLAS all-hadronic search~\cite{Aad:2012raa} is based on the full 2011 dataset which comprises 4.7 $\textrm{fb}^{-1}$ of $pp$ collision 
data at $\sqrt{s}$=7~TeV. Two top taggers optimised for different kinematic regimes are used: The \textit{HEPTopTagger}~\cite{Plehn:2009rk,Plehn:2010st} 
is applied to C-A $R=1.5$ jets with $p_T^{jet} > 200$~GeV and uses a combination of \textit{mass-drop tagging} and filtering to identify top jets. 
The \textit{TopTemplateTagger}~\cite{Almeida:2011aa} uses overlap functions that compare the energy flow in an anti-$k_T$ $R=1.0$ jet with the parton kinematics
in simulated templates of all-hadronic $t\bar{t}$ decays in order to quantify the resemblance of a jet with boosted top jets and jets initiated by a
single parton. The $p_T$ threshold for the leading (subleading) jet is 500~(450)~GeV. $Z'$ bosons are excluded at 95\% CL in the mass ranges
0.70-1.00~TeV as well as 1.28-1.32~TeV.

\begin{figure}
\centering
\subfloat[]{
  \includegraphics[width = 0.495\textwidth]{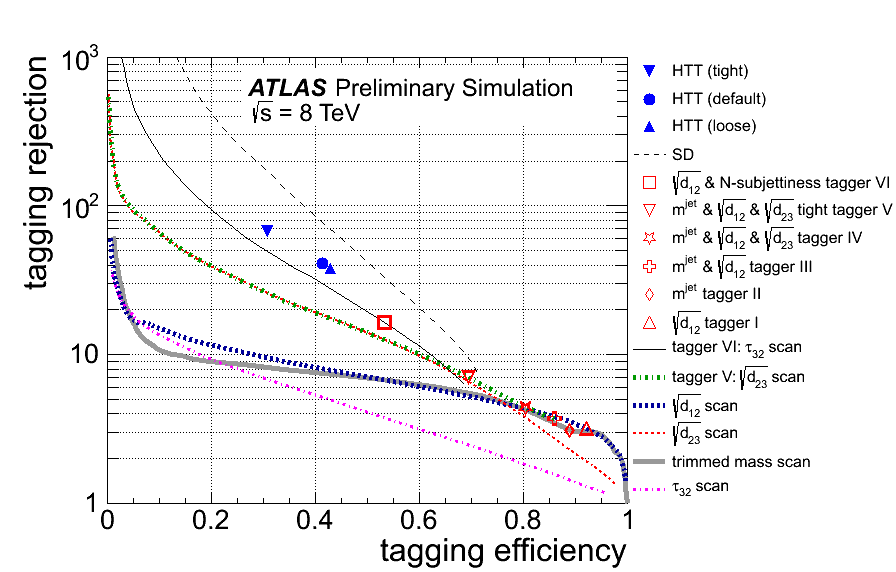}
}
\subfloat[]{
  \includegraphics[width = 0.415\textwidth]{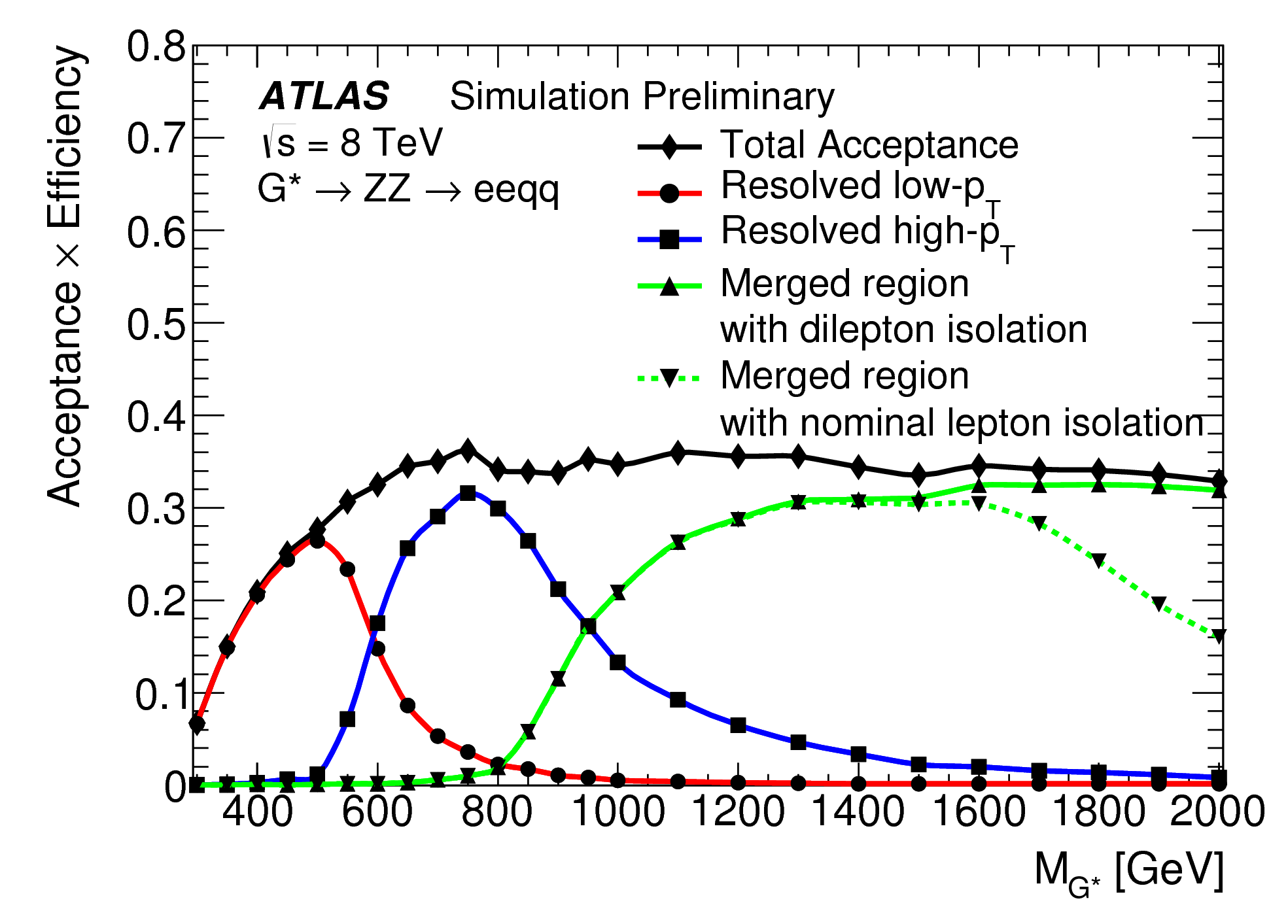}
}
\caption{ (a) Comparison of top taggers in terms of signal efficiency and rejection of multijet background~\cite{TheATLAScollaboration:2013qia}. 
Different final states require different top taggers, see text. (b) Contribution of the boosted and resolved channels to the overall
acceptance$\times$efficiency as a function of the ``bulk'' RS graviton mass $M_{G*}$ in the semileptonic diboson resonance search~\cite{Aad:2013myid}.}
\label{fig:Isolation}
\end{figure}

Figure~\ref{fig:Isolation}(a) shows a comparison of the performance of different top taggers in terms of signal efficiency 
and rejection of multijet background. The\textit{ HEPTopTagger} (blue markers) provides a high background rejection rate at the cost of a low signal 
efficiency, suited for all-hadronic searches. The lower background in the $1\ell+$jets search allows for a tagger with a higher signal
efficiency. The red cross marks the top tagger used in~\cite{ATLAS:2012bpa}.

\subsection{Searches for $W' \rightarrow tb$}

Both ATLAS and CMS have published searches for a heavy $W$ partner, $W'$, decaying via $W' \rightarrow tb$ with a hadronically decaying top quark
using 20 $\textrm{fb}^{-1}$ of $pp$ collision data at $\sqrt{s}$=8~TeV. 
The ATLAS search~\cite{Aad:2014xra} is optimised for $W'$ masses above 1.5~TeV where the boosted top quark is 
identified using a dedicated $W'$ \textit{Top Tagger}. This tagger uses trimmed anti-$k_T$ $R=1.0$ jets with $p_T^{jet} > 250$~GeV 
and cuts on the $k_T$ splitting scale $\sqrt{d_{12}}$ and two n-subjettiness variables to identify the 3-pronged substructure of a boosted top decay.
CMS~\cite{CMS:2014kfa} relies on the \textit{CMSTopTagger} with n-subjettiness as an additional identification criterion. 
Only jets with $p_T^{jet} > 450$~GeV are considered as top jet candidates.

% Both experiments set upper limits at 95\% CL on the production left- and right-handed $W'$ bosons as well as on the couplings to $tb$ as a function
% of $m_{W'}$ thus allowing for a model independent interpretation of the results:
ATLAS sets upper limits at 95\% CL on the $W' \rightarrow tb$ cross-section times branching ratio ranging between 0.16~pb and 0.33~pb 
for $W'$ bosons with purely lef-handed couplings, and between 0.10~pb and 0.21~pb for $W'$ bosons with purely right-handed couplings.
The sensitivity of the CMS search is further enhanced by combining the
results with those from the semileptonic channel where the top quark decays leptonically. This leads to an exclusion of right-handed $W'$ with masses
below 2.15~TeV at 95\% CL.
Both experiments also set upper limits at 95\% CL on the couplings to $tb$ as a function
of $m_{W'}$ thus allowing for a model independent interpretation of the results.

\subsection{Searches for stop quarks}

The application of boosted techniques in searches for supersymmetry is a relatively young field that has gained importance as the lower limits
on the masses of supersymmetric particles have been pushed towards the TeV-scale. Searches for stop quark pair production where boosted top quarks
and $W$ bosons arise from the decay $\tilde{t} \rightarrow t \chi_1^0$ have been published by both 
ATLAS (0- and 1-lepton channel~\cite{Aad:2014bva,Aad:2014bia}) and CMS (0-lepton channel~\cite{CMS:2013nia}).
For sake of brevity only the ATLAS 0-lepton search~\cite{Aad:2014bva} will be discussed here because it employs a novel substructure technique: 
\textit{jet reclustering}~\cite{Nachman:2014kla}. 
Here the large-$R$ jets are not built from calorimeter clusters or inner detector tracks but from small-$R$ jets. This particular search uses
anti-$k_t$ jets with $R=0.4$ and $p_T>$ 25~GeV as input to anti-$k_t$ clusterings with $R=0.8$ and 1.2. Here the $p_T$ cut on the small-$R$ 
jets acts like trimming. The event selection requires $\geq2$
reclustered $R=1.2$ jets that fulfill certain mass and $p_T^{jet,1.2}$ requirements. These are considered top jet candidates.
Another mass requirement is placed on the reclustered $R=0.8$ jet
with the highest $p_T^{jet,0.8}$ in order to suppress backgrounds without hadronic $W$ candidates. For a branching fraction of 100\% into $t \chi_1^0$
stop quark masses in the range 270-645~GeV are excluded at 95\% CL for $\chi_1^0$ masses below 30~GeV.

\section{Searches with Boosted Bosons}\label{sec:BoostedBosons}

Boosted techniques are becoming increasingly common in searches with hadronically decaying $W$, $Z$ and Higgs bosons in the final state.
Searches in all-hadronic final states especially benefit from substructure techniques as these allow for an effective control of the
large multi-jet background. Dedicated tagging techniques for boosted bosons have been studied by both ATLAS~\cite{ATL-PHYS-PUB-2014-004} 
and CMS~\cite{CMS-PAS-JME-13-006}.

CMS has published an inclusive search for resonances decaying to $qW$, $qZ$, $WW$, $WZ$ or $ZZ$ with fully hadronic boson decays~\cite{Khachatryan:2014hpa}.
In the boosted regime, the events exhibit a simple dijet topology. Signal events with boosted bosons are tagged by 
requiring one or two pruned C-A jets with $m^{jet}$ between 70 and 100~GeV and two-pronged substructure identified using the n-subjettiness variable. 
The lower mass limits on excited quarks, RS1 gravitons and $W'$ bosons, set using 19.7 $\textrm{fb}^{-1}$ of $\sqrt{s}$=8~TeV data, 
are all above 1~TeV hence accounting for the presence of boosted bosons.

Resonance searches in the semileptonic final state have been conducted by both ATLAS~\cite{Aad:2013myid} and CMS~\cite{Khachatryan:2014gha}. 
The ATLAS search focuses on decays to $WZ$ and $ZZ$ with at least one leptonic decay $Z \rightarrow \ell^+\ell^-$. 
Both resolved and boosted scenarios are considered. In the boosted case, modified lepton isolation criteria are used
to maintain a high selection efficiency for events where the two leptons from the boosted $Z$ decay get into each others isolation cones. 
Boosted hadronic boson decays are identified by applying a slightly modified version of the \textit{mass-drop-filtering} technique from 
Ref~\cite{Butterworth:2008iy} to C-A $R=1.2$ jets. 
The same boson tagger is used in a search for dark matter with a single boosted boson in the final state~\cite{Aad:2013oja}. The combination of 
resolved and boosted techniques, including the modified lepton isolation, provides a stable signal efficiency over the whole range
of potential resonance masses, as illustrated in Figure~\ref{fig:Isolation}(b).

A search involving boosted Higgs bosons has been conducted by CMS~\cite{CMS:2014afa}.
It looks for pair production of vector-like $B'$ quarks which are predicted by many models involving top partners and decay via $B' \rightarrow Hb$.
The search has been optimised for boosted $H\rightarrow b\bar{b}$ decays, the dominant Higgs decay mode with 56\% branching ratio. 
In this difficult final state, substructure techniques are a major asset in reducing the multi-jet background. Higgs jets are 
reconstructed as pruned C-A $R=0.8$ jets passing a cut on 2-subjettiness and b-tagging requirements on both subjets. 
The search excludes $B'$ quarks for masses below 846~GeV at 95\% CL based on 19.7 $\textrm{fb}^{-1}$ of $\sqrt{s}$=8~TeV data.

\section{Searches for Top Quark Partners}\label{sec:TopPartners}

Top quark partners play an important role in many extensions of the SM since they allow for cancellation of the quadratically divergent
quantum-loop corrections to the Higgs boson mass introduced (predominantly) by the top quark. Their decays usually involve both top quarks
and heavy gauge or Higgs bosons in the final state which are boosted if the hypothetical top partner is sufficiently heavy.

\begin{figure}
\centering
\subfloat[]{
  \includegraphics[width = 0.435\textwidth]{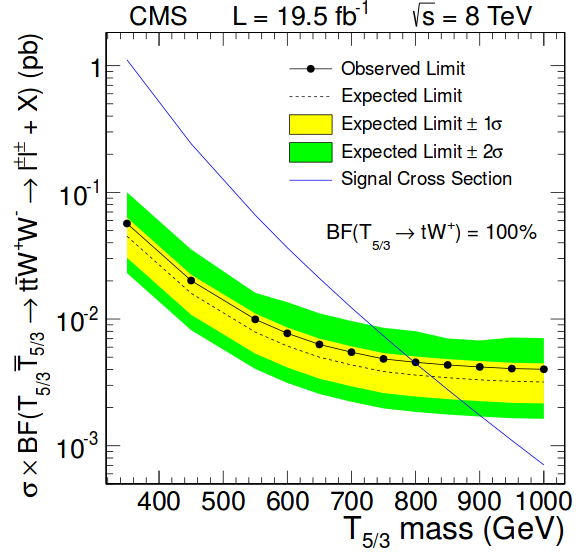}
}
\subfloat[]{
  \includegraphics[width = 0.505\textwidth]{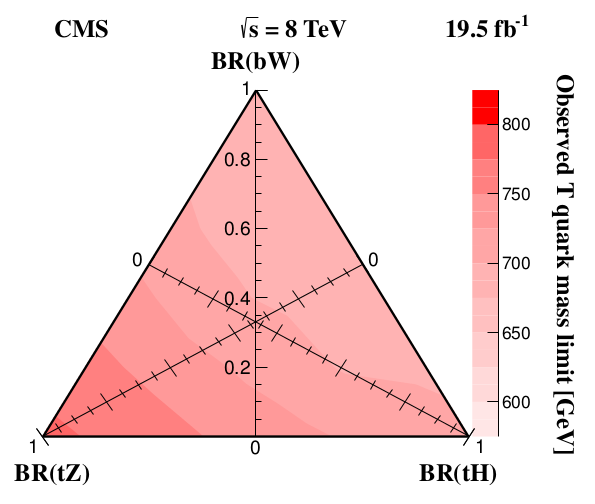}
}
\caption{(a) Upper cross-section limit as a function of $m_{T_{5/3}}$~\cite{Chatrchyan:2013wfa}. 
(b) Branching-fraction triangle with observed 95\% CL limits on
the mass of a $2e/3$ $T$ quark~\cite{Chatrchyan:2013uxa}.
}
\label{fig:TopPartners}
\end{figure}

A search for pair production of a vector-like top partner with charge $\pm 5e/3$, $T_{5/3}$,
has been conducted by CMS~\cite{Chatrchyan:2013wfa} using 19.5 $\textrm{fb}^{-1}$ of $\sqrt{s}$=8~TeV $pp$ collision data. The top partner is expected
to decay via $T_{5/3} \rightarrow t W$, and the search focuses on the same-sign dilepton final state where the top quark and $W$ boson from at least
one $T_{5/3}$ both decay leptonically. Boosted top quarks are identified via the \textit{CMSTopTagger} while boosted hadronic $W$ bosons
are reconstructed as pruned C-A $R=0.8$ jets with exactly two subjets and a mass compatible with $m_W$. $T_{5/3}$ masses below 800~GeV are excluded
at 95\% CL as illustrated in Figure~\ref{fig:TopPartners}(a).

Several searches have been conducted for pair production of vector-like quarks, $T$, with the same charge as the top quark. A $T$ quark can decay into
three final states: $Wb$, $Zt$ and $Ht$ where the branching ratios are model dependent. CMS has conducted the first inclusive search~\cite{Chatrchyan:2013uxa} for all three
decay modes in final states with at least one isolated lepton using the 19.7 $\textrm{fb}^{-1}$ of $\sqrt{s}$=8~TeV data. 
Boosted hadronic top quarks are tagged using the \textit{CMSTopTagger}, and boosted $W$ bosons are reconstructed as pruned C-A $R=0.8$ jets 
with a mass close to $m_W$. Mass limits are set as a function of the branching ratios as illustrated
in Figure~\ref{fig:TopPartners}(b). A search in the same final state but assuming a 100\% branching ratio into $Wb$ has been conducted by 
ATLAS~\cite{ATLAS:2012qe}. Boosted hadronic $W$ decays are reconstructed as anti-$k_T$ $R=0.4$ jets with a mass close to $m_W$ and mini-isolation
is required on muons. Searches optimised for the $Ht$ decay mode with $H \rightarrow b\bar{b}$ have been published by both experiments. 
Only the CMS search~\cite{CMS:2014aka} uses boosted techniques and will be discussed here.
Both boosted top quarks and boosted Higgs bosons are reconstructed from filtered C-A $R=1.5$ jets with $p_T>$150~GeV. 
The top jet is identified using the \textit{HEPTopTagger} whereas the Higgs jet is required to have at least two $b$-tagged $R=0.3$ subjets 
with invariant dijet mass greater than 60~GeV.
Assuming a 100\% branching ratio to $Ht$, the observed limit
on $m_{T}$ at 95\% CL is 747~GeV based on 19.7~$\textrm{fb}^{-1}$ of $\sqrt{s}$=8~TeV data.

\section{Summary}
Boosted objects are key elements in searches for new physics at the high energy and mass scales accessible at the LHC because they provide
sensitivity in kinematic regimes where traditional reconstruction techniques fail. No deviation from the SM has been observed during Run-I
and upper limits on many benchmark models have been pushed into the TeV regime. This together with the planned increase of the center-of-mass energy
of the LHC to 13~(later 14)~TeV from 2015 onwards will further boost the number of searches (as well as measurements) relying on these novel techniques.
The era of boosted objects has only just begun.

%%%%%%%%%%%%%%%%%%%%%%%%%%%%%%%%%%%%%%%%%%%%%%%%%%%%%%%%%%%%%%%%%%%%%%%%%
%%
%%   use this format to include a figure into your paper.   Please use .pdf files for pdflatex compilation.
%%
% \begin{figure}[htb]
% \begin{minipage}{0.49\textwidth}
%  \centering
%  \includegraphics[width = 0.48\textwidth]{figures/dR_b_W_ptt.png}
% \end{minipage}
% \begin{minipage}{0.49\textwidth}
%  \centering
%  \includegraphics[width = 0.48\textwidth]{figures/CMS_Hbb_dR.png}
% \end{minipage}
% \caption{}
% \label{fig:magnet}
% \end{figure}
%%%%%%%%%%%%%%%%%%%%%%%%%%%%%%%%%%%%%%%%%%%%%%%%%%%%%%%%%%%%%%%%%%%%%%%%%%%

%%%%%%%%%%%%%%%%%%%%%%%%%%%%%%%%%%%%%%%%%%%%%%%%%%%%%%%%%%%%%%%%%%%%%%%%%
%%
%%   use this format to include a LaTeX table  into your paper
%%
% \begin{table}[t]
% \begin{center}
% \begin{tabular}{l|ccc}  
% 
% \end{center}
% \end{table}
%%%%%%%%%%%%%%%%%%%%%%%%%%%%%%%%%%%%%%%%%%%%%%%%%%%%%%%%%%%%%%%%%%%%%%%%%%%

\Acknowledgements

% \bibliographystyle{is-unsrt}
% \bibliography{ProceedingsPIC2014.bib}

% \small

\end{document}